\documentclass[aip,reprint]{revtex4-1}

\usepackage{graphicx}
\PassOptionsToPackage{caption=false}{subfig}
\usepackage{subfig}
\usepackage{hyperref}

\def\etal{{\em et al.}}

\begin{document}

\title{A readout for large arrays of Microwave Kinetic Inductance Detectors}

\author{Sean McHugh}
\email{mchugh@physics.ucsb.edu}
\affiliation{Department of Physics, University of California, Santa Barbara, CA 93106}

\author{Benjamin A. Mazin}
\affiliation{Department of Physics, University of California, Santa Barbara, CA 93106}

\author{Bruno Serfass}
\affiliation{Department of Physics, University of California, Berkeley, CA 94720}

\author{Seth Meeker}
\affiliation{Department of Physics, University of California, Santa Barbara, CA 93106}

\author{Kieran O'Brien}
\affiliation{Department of Physics, University of California, Santa Barbara, CA 93106}

\author{Ran Duan}
\affiliation{Department of Physics, California Institute of Technology, Pasadena, CA 91125}

\author{Rick Raffanti}
\affiliation{Techne Instruments, 4066 Oakmore Rd, Oakland, CA 94602}

\author{Dan Werthimer}
\affiliation{Department of Physics, University of California, Berkeley, CA 94720}

\date{\today}

\begin{abstract}
Microwave Kinetic Inductance Detectors (MKIDs) are superconducting detectors capable of counting single photons and measuring their energy in the UV, optical, and near-IR.  MKIDs feature intrinsic frequency domain multiplexing (FDM) at microwave frequencies, allowing the construction and readout of large arrays.  Due to the microwave FDM, MKIDs do not require the complex cryogenic multiplexing electronics used for similar detectors, such as Transition Edge Sensors (TESs), but instead transfer this complexity to room temperature electronics where they present a formidable signal processing challenge.  
In this paper we describe the first successful effort to build a readout for a photon counting optical/near-IR astronomical instrument, the ARray Camera for Optical to Near-infrared Spectrophotometry (ARCONS).  This readout is based on open source hardware developed by the Collaboration for Astronomy Signal Processing and Electronics Research (CASPER).  Designed principally for radio telescope backends, it is flexible enough to be used for a variety of signal processing applications.
\end{abstract}


\maketitle

\section{Introduction}
ARCONS, the ARray Camera for Optical to Near-infrared Spectrophotometry, was successfully commissioned at the Coud\'{e} focus of the Palomar 200 inch telescope on July 28, 2011.  ARCONS is based on a 32$\times$32 array of Microwave Kinetic Inductance Detectors (MKIDs), which are photon-counting detectors that have an intrinsic energy resolution R$\sim$20--150, enabling them to perform low-resolution spectroimaging without filters or dispersive optics~\cite{mazinoe11}.  ARCONS was designed to be sensitive in the 0.4-1.1~$\mu m$ wavelength range and can time-tag photons to 1~$\mu$s without read noise or dark current.  ARCONS is the first astronomical instrument based on optical/near-IR MKIDs, and is the largest non-dispersive optical/near-IR spectrophotometer fielded by a factor of 10.  More details on the instrument design and goals can be found in Mazin \etal~\cite{Mazin:2010p3814}.  

\begin{figure}
\begin{center}
\includegraphics[width=3.5in]{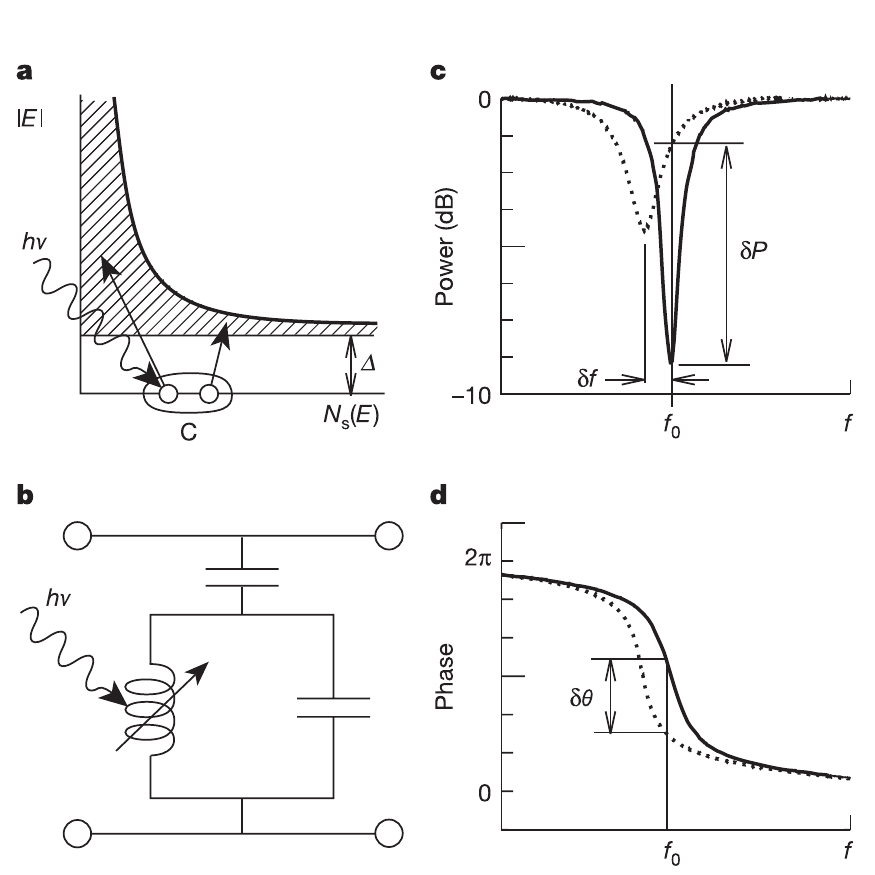}
\end{center}
\caption{The basic operation of an MKID, republished from Day \etal\cite{day03}. (a) Photons with energy $h\nu$ are absorbed in a superconducting film, producing a number of excitations, called quasiparticles.  (b) To sensitively measure these quasiparticles, the film is placed in a high frequency planar resonant circuit.  The change in the surface impedance of the film following a photon absorption event pushes the resonance to lower frequency and changes the amplitude of the transmission through the circuit.  The dotted line in (c) shows the change of the transmitted microwave signal due to an incident photon.  To monitor the resonant circuit, it is continuously excited with a microwave signal.  (d) The phase of the transmitted microwave signal depends on the proximity of the continuous microwave signal frequency and the variable resonant circuit frequency.  The energy of an absorbed photon can be determined by measuring the degree of phase shift of the transmitted microwave signal.} \label{fig:detcartoon}
\end{figure}

MKIDs work on the principle that incident photons change the surface impedance of a superconductor through the kinetic inductance effect~\cite{mattis58}.  The kinetic inductance effect occurs because energy can be stored in the supercurrent of a superconductor.  Reversing the direction of the supercurrent requires extracting the kinetic energy stored in the supercurrent, which yields an extra inductance. This change can be accurately measured by placing this superconducting inductor in a lithographed resonator.  A microwave probe signal is tuned to the resonant frequency of the resonator, and any photons which are absorbed in the inductor will imprint their signature as changes in phase and amplitude of the probe signal.  Since the measured quality factor of the resonators is high and their microwave transmission off resonance is nearly perfect, multiplexing can be accomplished by tuning each pixel to a different resonant frequency with lithography during device fabrication.  A comb of probe signals (one frequency for each resonator) can be sent into the detector array, and room temperature electronics can recover the changes in amplitude and phase without significant crosstalk~\cite{day03}, as shown in Figure~\ref{fig:detcartoon}.

Low temperature operation places serious constraints on any cryogenic detector.  The obvious advantage of MKIDs over competing cryogenic technologies like Transition Edge Sensors~\cite{irwin96} (TESs) is the elimination of the cryogenic electronics required for multiplexing~\cite{chervenak99,yoon01}. For ARCONS, only two microwave feed lines are used to read out 1024 MKIDs operating at 100 mK.  However, a trade-off exists between the MKID's cryogenic simplicity and the complexity of the room temperature electronics.  In particular, the signal processing task of deciphering the complicated sum of all resonator probe signals is not unlike the challenge presented to modern communication engineers when developing multi-channel cell base transceiver stations.  In this paper, we discuss in detail the MKID readout developed for ARCONS.  Although various aspects are unique to ARCONS, the readout is broadly applicable to any other MKIDs or MSQUID~\cite{Niemack:2010p4657} application, such as submillimeter cameras~\cite{Monfardini:2010p6990,Maloney:2010p6993} or dark matter detectors~\cite{Golwala:2008p3970}.  Our approach differs from previous FDM MKID readouts~\cite{Mazin:2006p5,Yates:2009p3223,Bourrion:2011p7015} in that it uses a powerful and flexible channelization technique (Section~\ref{sec:channelizer}) better suited to the high speeds needed for near-IR/Optical/X-ray photon detection instead of a long, single stage Fourier Transform to discriminate between readout tones.  This readout also and covers significantly more bandwidth than previous efforts.

\section{Hardware}
\subsection{Overall Approach}

\begin{figure*}
\begin{center}
\includegraphics[width=7.0in]{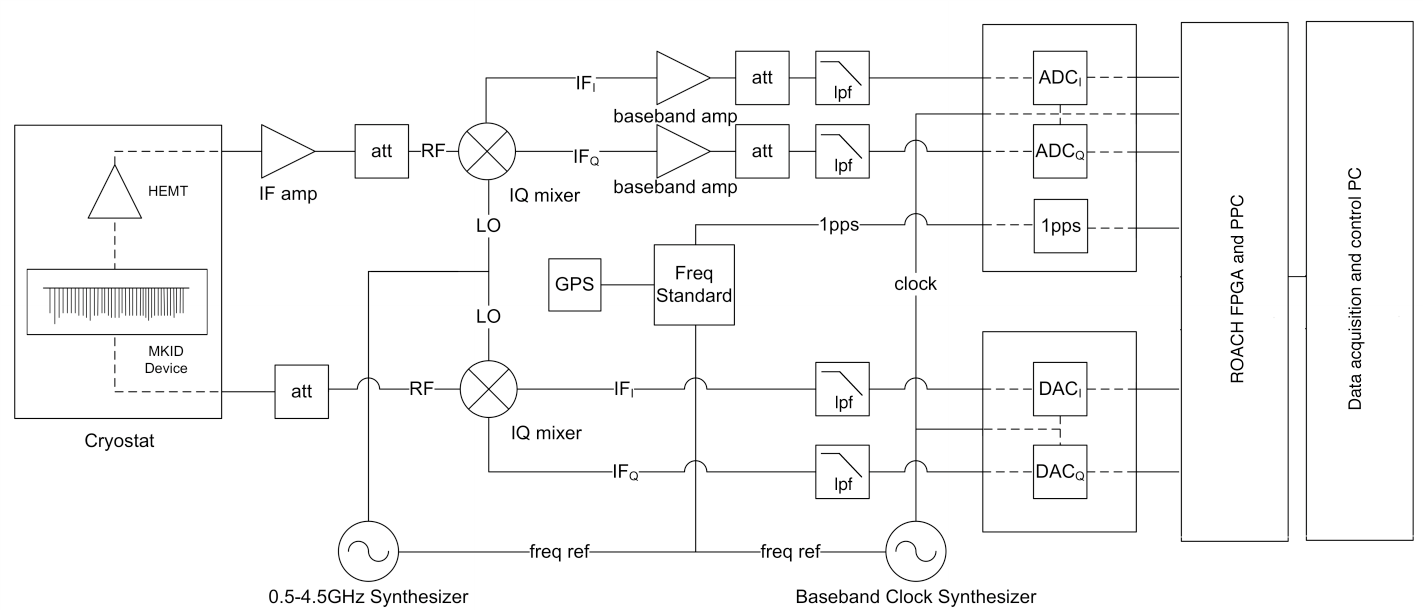}
\end{center}
\caption{The block diagram of the SDR readout for ARCONS.} \label{fig:sdr}
\end{figure*}

In order to read out an array of MKIDs, one must generate probe signals at the resonant frequency of each MKID.  The sum of these waveforms, or ``frequency comb," is sent through the device, where each detector imprints a record of its illumination on its corresponding probe signal.  The frequency comb is then amplified with a cryogenic HEMT and brought outside the cryostat, where it is digitized and the phase and amplitude modulation of each individual probe signal is recorded with room temperature electronics.  

The MKID readout we developed to perform these operations is based on the CASPER's Reconfigurable Open Architecture Computing Hardware (ROACH) board~\cite{parsons_et_al2006}.  The core of the ROACH board is a Xilinx Virtex 5 SX95T field programmable gate array (FPGA) used for demanding signal processing operations.  In addition the ROACH board accommodates external hardware to generate and digitize analog signals required to read out MKIDs.  However, the MKID resonant frequencies are above the range accessible to common digital-to-analog (DAC) and analog-to-digital (ADC) converters.  Therefore, an intermediate step must be taken to mix the ``baseband" signals, which can be processed by the ADC and DAC, to the higher frequencies at which the the MKIDs resonate.  To do this, we employ the quadrature amplitude modulation strategy generally employed in the digital communications field: A local oscillator (LO) is used to generate a continuous microwave tone at frequency, $f_{LO}$, near the resonators' frequencies.  A baseband signal is precomputed and output from a two-channel DAC.  For each resonator, the baseband signal is composed of two waveforms: a sine wave at the resonant frequency minus $f_{LO}$ ($I_{BB}$), and the same waveform shifted -90 degrees out-of-phase ($Q_{BB}$).  This complex baseband signal then multiplied by the the complex LO to produce a probe signal at the desired MKID frequency, $I_{BB}\mbox{cos}(2\pi ft) + Q_{BB}\mbox{sin}(2\pi ft)$.  After passing through the device and amplifiers the probe signal is mixed back down to the baseband with the same LO where it can be digitized with a two-channel ADC.  Using this method with two-channel ADCs and DACs allows us to double the bandwidth available to a single DAC, and probe MKIDs above and below the $f_{LO}$.

After digitization, the signals are passed to the FPGA since the data rate is far too high to store for offline processing.  Once in the FPGA, the essential task is to break down the broadband signal emerging from the MKIDs, parsing it into the narrowband frequency bins constituting each channel in a process known as ``channelizing."  We are using a two stage channelization core that uses a polyphase filter bank (PFB)\cite{DSP} followed by a time-multiplexed direct digital downconverter to allow us to readout 256 resonators in 512 MHz of bandwidth, as discussed in Section~\ref{sec:channelizer}.  A schematic for the readout hardware is shown in Figure~\ref{fig:sdr}.

\subsection{Design Requirements}
The requirements for a MKID readout are relatively straightforward.
\begin{itemize}
\item The readout must not introduce significant noise above the system noise floor set by the cryogenic HEMT amplifier with a noise temperature of $\sim$6 K.
\item For ARCONS, the entire readout system must be capable of reading out 1024 resonators in $\sim$2 GHz of bandwidth.
\item The output bandwidth of the readout must be wide enough to faithfully reconstruct the desired data.  
\item Crosstalk between channels greater than 250 kHz apart should be less than 1\%.
\end{itemize}

Conventional electronic phase noise is measured with respect to the origin of the $IQ$ plane.  Although it is often convenient to characterize a MKID's phase noise with respect to the center of the resonator's $IQ$ loop, we maintain the more general convention here.  All mentions of phase noise are calculated and measured with respect to the origin of the $IQ$ plane.  The double sideband (DSB) phase noise of the HEMT is given by the simple expression, $k_BT_n/P$, where $T_n$ is noise temperature of the amplifier, and $P$ is the power on the input.  For ARCONS, the readout power for each MKID is roughly $-100 \pm15$ dBm.  For -85 dBm readout power, off resonance, and $T_n = 6$ K, we expect -106 dBc/Hz for the HEMT phase noise.  Ideally, the readout electronics will contribute noise well below this value.  

\begin{figure}
\begin{center}
\includegraphics[width=3.5in]{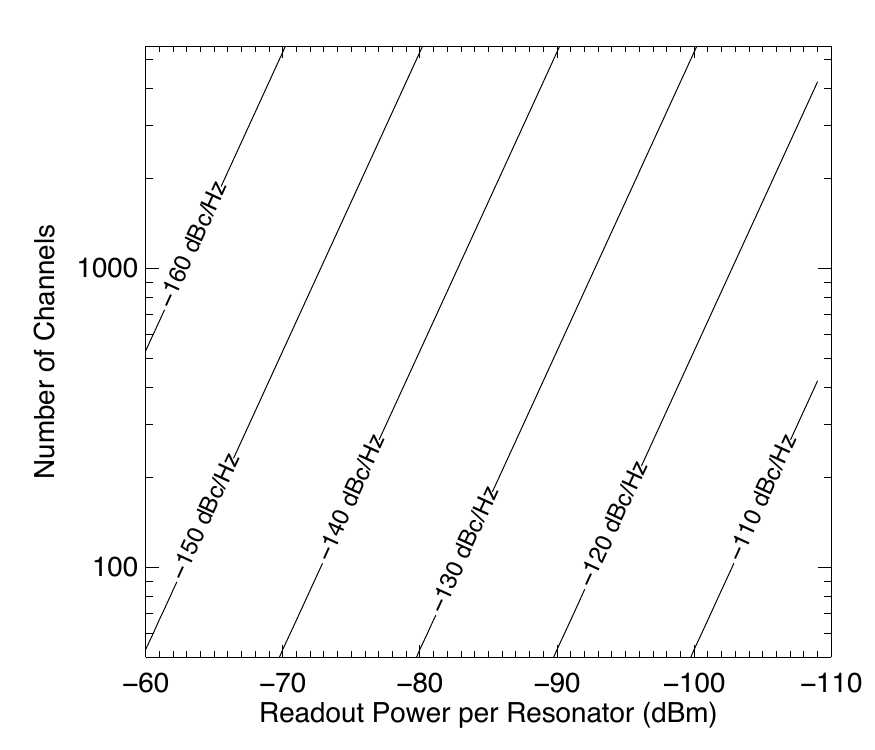}
\end{center}
\caption{The required noise floor density in dBc/Hz of an ADC as a function of the number of resonators being read out and the readout power per resonator for an amplifier noise temperature of 6 K.  With a 512 MSample/s sample rate, 10.6 ENOB, we expect about -147 dBc/Hz for the ADC's phase noise.  Even at the relatively high readout power -85 dBm this would allow for reading out over 1000 channels.  Alternatively, 256 channels could be read out with power as high as -70 dBm.} 
\label{fig:PvNc}
\end{figure}

As the readout power is increased, the phase noise of the HEMT decreases placing stricter requirements on the readout electronics.  Given an ADC's voltage noise specification, the number of channels $n$ an ADC can read out at a given power without adding to the HEMT noise can be calculated.  The results are shown in Figure~\ref{fig:PvNc}.  The contours of Fig.~\ref{fig:PvNc} are calculated by scaling the HEMT noise with a channel-dependent factor giving, $k_BT_n/(p^2_{max}P)$, and assuming a 3 dB margin between the HEMT and the ADC noise.  $p_{max}$ sets the full scale amplitude achievable by each tone when digitizing a multi-tone waveform.  Generally, $p_{max}$ is minimized by assuming the phase of each tone is random resulting in $p_{max} \propto n^{1/2}$ for $n>>1$.  Specifically, we find that  $p_{max} = 2.75\times n^{1/2}$ provides a reasonable estimate and was assumed when calculating Figure~\ref{fig:PvNc}.  (See Section V. for further discussion.) 

The 1024 detectors in the ARCONS are contained in two arrays on a single wafer, with each array accessed by a single microwave feed line also fabricated on the surface of the wafer.  Each array contains 512 resonators separated by approximately 2 MHz occupying roughly 1 GHz bandwidth above 4 GHz.  As discussed below, a single readout board can read out 256 resonators in up to 512 MHz of bandwidth, meaning that two readout boards, multiplexed with a power combiner and power splitter, are required for each feedline on the MKID array.  A total of four boards is then required for the complete readout.

Finally, we require a 500 kHz output sample rate for each resonator, which allows us to capture the pulse profile generated by the photon faithfully, yet minimizing crosstalk between resonators with close resonant frequencies.  
	
\subsection{IF Board}
Since the MKID resonant frequencies are above the range accessible to the ADC and DAC, which can only record or generate Nyquist-limited signals $\le 256$ MHz, we must mix the base band signals, with a microwave local oscillator near the resonator frequency range.  The product of the LO and the two base band signals, called the intermediate frequency (IF), can then be used to probe MKIDs within $\pm$256 MHz of the LO frequency.  We have constructed a circuit board containing two Analog Devices ADF4350 wideband synthesizer, which can be programmed to generate an output frequency between 137.5 MHz and 4.4 GHz.  One synthesizer provides the LO for mixing and the other provides the 512 MHz logic clock for the ADC, DAC, and FPGA.  In addition, this IF board contains additional room temperature hardware required for the readout: two IQ mixers for mixing to and from the IF; three IF wideband amplifiers; one baseband amplifier; three programmable attenuators; and a frequency doubler used to generate LO frequencies above 4.4 GHz.  The key figure of merit for the IQ mixers is the sideband suppression, which we measure to be -30 dBc.  Physically, the IF board connects to the DAC and ADC boards via SMA connectors, and provides IF out and in SMA connectors which lead to the cryostat housing the MKIDs.

\begin{figure}
\begin{center}
\includegraphics[width=3.5in]{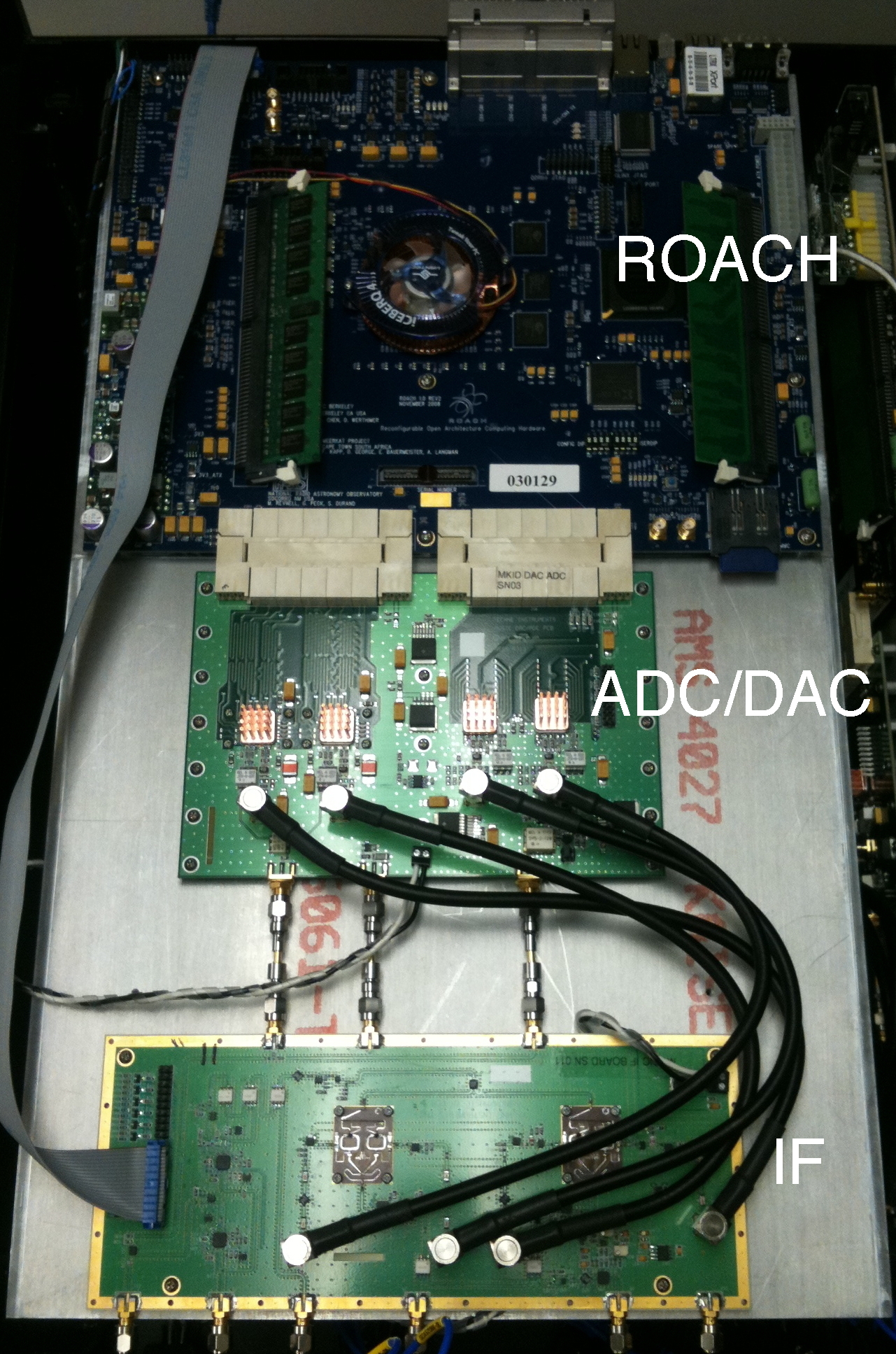}
\end{center}
\caption{One of four ROACH assemblies including the ADC, DAC, and IF boards.}
\label{fig:crate2}
\end{figure}

\subsection{ADC/DAC Boards}
The two pieces of hardware that place the most stringent limits on the readout electronics are the DAC for generating the probe signals, and the ADC for digitizing the signal for signal processing.  As discussed above, the signal to noise ratio (SNR), and sample rate are the crucial figures of merit.  In general, ADCs underperform DACs and therefore place the strictest limitations.  We selected a 12-bit, 550 MSample/s ADC from Texas Instruments (ADS5463), which has a SNR of approximately 64 dB and an effective number of bits (ENOB) of 10.6 for a full scale input across the bandwidth.  The TI ADS5463 has a noise floor density $ -20\log(2^{10.6}) - 10 \log(N_s/2)$ of -147 dBc/Hz, where $N_s$ is the ADC sample rate.  According to Figure~\ref{fig:PvNc}, this allows reading out over 1000 channels per board, even at the relatively high readout power of -85 dBm.  Alternatively, 256 channels could be read out with power as high as -70 dBm before ADC noise becomes comparable to the HEMT noise.

For the DAC, we selected the 16-bit 1 GSample/s DAC (DAC5681) from Texas Instruments, which has a SNR of roughly 75 dB and noise floor of -159 dBc/Hz; again, well below the SNR requirements imposed by the HEMT, shown in Figure~\ref{fig:PvNc}.  

The baseband signals are generated by two DACs, one for each I and Q, on a custom board shared with the ADCs.  After the IF signal is passed through the MKIDs and mixed down to the baseband, it is recorded by two ADCs.  The ADC/DAC boards mate directly with the peripheral ports of the ROACH board, and the SMA connectors on the IF board.  The DAC board also routes the logic clock generated on the IF board to the FPGA, divided by a factor of two.  The ADC board has an additional SMA input for routing a synchronization signal via the IF board to the FPGA.  This was used to route a one pulse-per-second (PPS) timing signal from an external GPS.  For ARCONS, the DAC and ADC were clocked at 512 MHz, and the FPGA at 256 MHz.  The SNR of both ADC and DAC boards are within 3 to 5 dB of the specifications in the data sheets.

\subsection{Complete Readout}
\begin{figure}
\begin{center}
\includegraphics[width=3.5in]{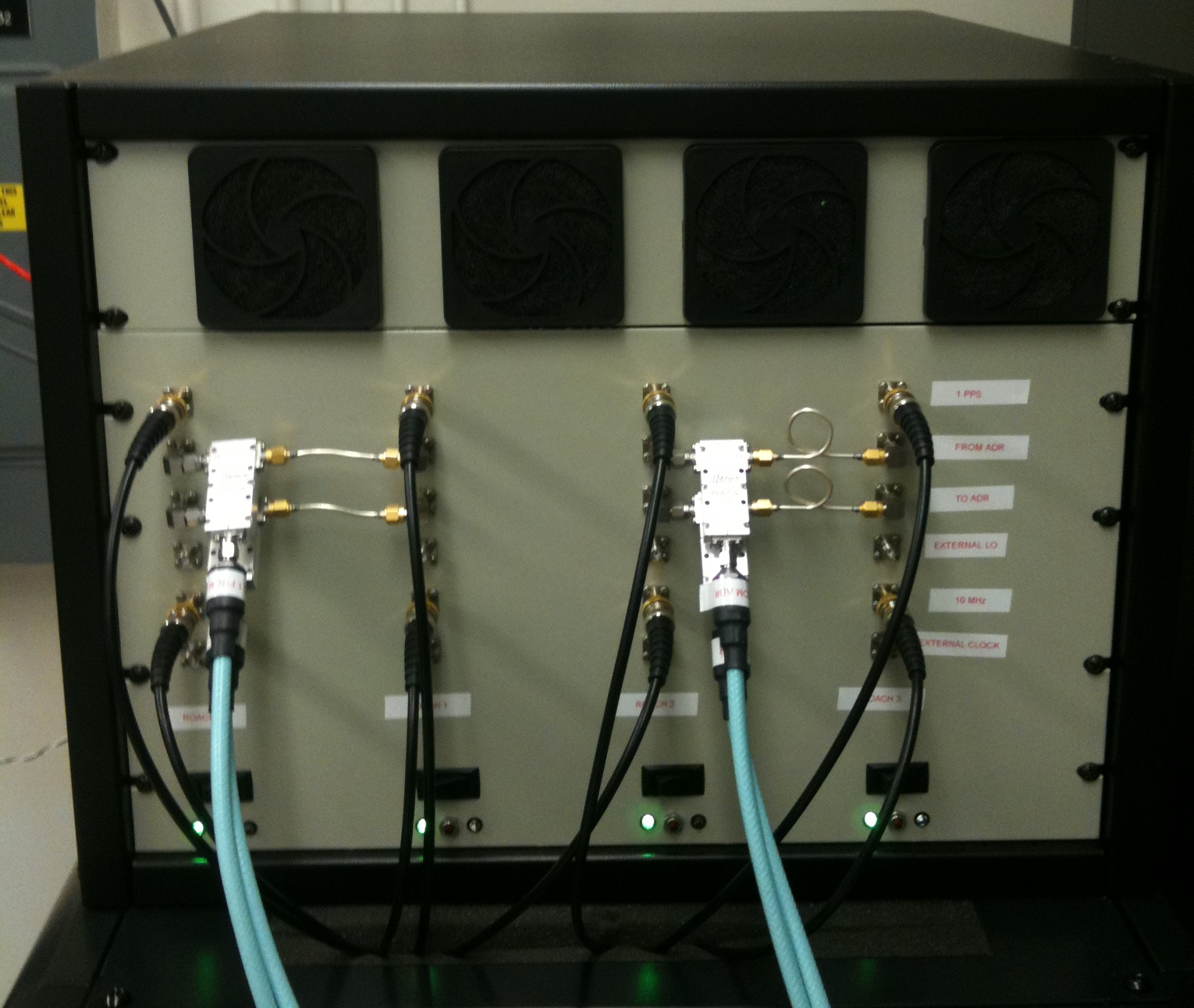}
\end{center}
\caption{Custom crate to house up to eight ROACH assemblies.  The SMA connectors to each IF board are visible on the front panel.}
\label{fig:crate1}
\end{figure}

We have constructed a custom crate to house the four ROACH boards required to readout the entire MKID array for a compact installation at the observatory.  It can be upgraded to house eight boards.  The crate is pictured in Figure~\ref{fig:crate1}.  Within the crate are four separate assemblies containing a ROACH, ADC/DAC, and IF boards, as pictured in Figure~\ref{fig:crate2}.

There are two additional components that make the readout complete, and are essential to the accurate estimation of the photon arrival times: a Meinberg (GPS170PEX) GPS clock mounted within the data acquisition PC, and a Stanford Research (FS725) rubidium 10 MHz frequency standard.  The GPS generates a 1 PPS signal to which the rubidium frequency standard is synchronized, and the 10 MHz signal is used by both the IF and clock synthesizers.  With these, we are able to record the time with better than 1$\,\mu$s resolution and maintain synchronization between the four ROACH boards.

\section{Firmware}

\subsection{Comb Generation}

The most computationally intensive signal processing tasks are performed on the FPGA as firmware, including the generation of the input signal comb.  Rather than directly generating a waveform for each resonant frequency on the FPGA, a waveform defined as the sum of all resonant frequencies is loaded into a portion of the DRAM memory.  Each memory location contains data accessed sequentially as a look-up table (LUT) and output to the DAC.  The FPGA is clocked at 256 MHz, while the ADC and DAC are clocked at 512 MHz.  Therefore, in order to generate the waveform output for each DAC channel $I$ and $Q$, two consecutive 16 bit time samples must be presented to the DAC at each FPGA clock cycle. 

\begin{figure*}
\includegraphics[width=7.0in]{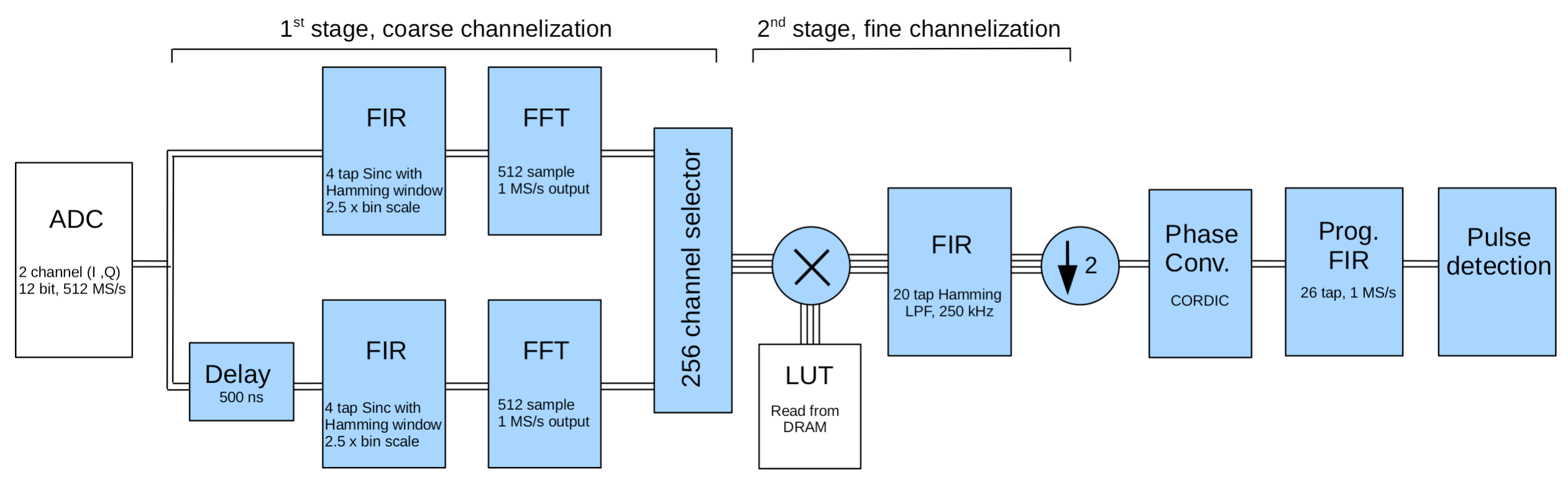}
\caption{Schematic of channelizer firmware.}
\label{fig:chan}
\end{figure*}

\subsection{Channelizer}
\label{sec:channelizer}
The remainder of the FPGA resources are dedicated to processing the data digitized by the ADC.  The first step and most mathematically involved process is the channelization.  By design, the resonant frequencies of the MKIDs are evenly spaced about 2 MHz apart.  The designed quality factor of the MKIDs and physics of the superconducting material sets the channelizer bandwidth for each resonator at 250 kHz.  Variations in the fabrication are unavoidable, however, and the actual spacing of the resonant frequencies can vary considerably.  This constraint means the channelizer must be able to accommodate 500 kHz wide (double side band) channels arbitrarily spaced in the 512 MHz bandwidth available to each ROACH board.  Digital down conversion (DDC) is the most straight-forward and common solution to this task.  However, the hardware demands of this scheme are high.  We estimate that the number of channels processed per FPGA can be a factor of at least 10 higher with the more sophisticated, albeit complicated, scheme shown in Figure~\ref{fig:chan}. 


\begin{figure*}
\includegraphics[width=7.0in]{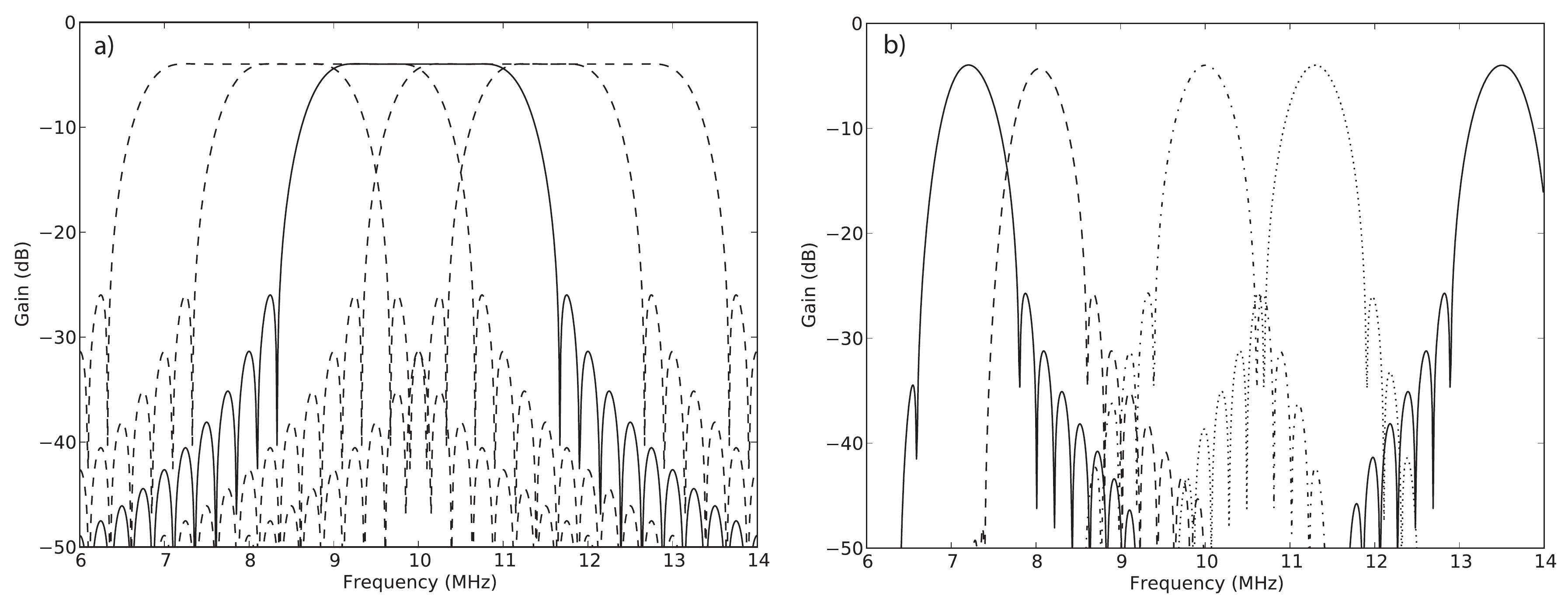}
\caption{(a) Three sample frequency bins output from the 1st stage of channelization with the PFB.  Each bin is 2 MHz wide, with $\sim$1 MHz overlap with adjacent bins.  (b) The irregularly spaced, narrow frequency bins output from the 2nd stage.  The bin width is 500 kHz wide and centered on the resonant frequency for a single pixel.}
\label{fig:pfb}
\end{figure*}

We perform the channelization in two stages.  The first is a coarse channelization resulting in wide, evenly spaced, overlapping bins with frequency widths equal to twice the frequency spacing.  Figure~\ref{fig:pfb}(a) shows an example of the overlapping bins output from the first stage of channelization.  This coarse channelization is done with a twice over-sampled Fast Fourier Transform (FFT) preceded by a finite impulse response filter (FIR).  The combined effect is to yield a Polyphase Filter Bank (PFB)\cite{DSP}. The time multiplexed frequency bins of the $L=512$ samples FFT are output from the first stage.  With the FPGA clock running at 256 MHz, two samples per frequency bin per clock cycle are output from the PFB.  Since a FFT has a sinc-like frequency response, the time stream is first conditioned with the FIR$_{PFB}$.  This FIR$_{PFB}$ is a FIR filter with a set of coefficients drawn from a sinc function.  Ideally, this convolution with a sinc function in the time domain results in rectangular frequencies output from the FFTs.  More taps in the FIR$_{PFB}$ gives more rectangular frequency bin widths, however, this also consumes significant FPGA resources that could be used in later stages of the firmware.  In order to produce the oversampled 2 MSample/s/channel output, two copies of the digitized time stream from the ADC are sent to two FIR$_{PFB}$s and FFTs to be processed in parallel, with one copy delayed by $L/2$.  In addition, the odd frequencies output from the non-delayed block are multiplied by $-1$ in order to maintain consistent phases of the output between the two FFTs.  The output of these two FFTs are then treated as consecutive time samples.   See the appendix for further mathematical details on the PFB.

The second stage is a fine channelization done with the output of the first stage with DDC.  The 256 channels are presented at the output of the first stage in a time multiplexed fashion with a sample rate of 2 MSample/s (2 samples/FPGA clock cycle).  Unless the probe signal is exactly on a frequency bin center of the first stage, the output will beat with a frequency that is the difference between the probe signal for a particular resonator and the nearest bin center frequency from the first stage.  This is also the frequency with which the signal is digitally down-converted.  Rather than using the conventional direct digital synthesis of this signal, we pre-calculate it as a LUT and store it in the ROACH board's DRAM.  This is efficient and convenient since the DAC LUT and the DDC LUT each use 64 bits of the 128 bit DRAM-to-FPGA data bus width.  The DDC LUT from the DRAM is presented in a time multiplexed fashion to match the channels' output from the first stage.  

The complex signal ($I, Q$) emerging from the mixing stage must now be filtered with a low-pass filter.  As stated above, we choose a 20-tap, Hamming filter with a 250 kHz cutoff frequency.  In addition, this output is down-sampled by a factor of two.  Figure~\ref{fig:pfb}(b) shows an example of the fine channelization.  The output from the DDC could be downsampled an additional factor of two at this point, however for ease of implementation this was not done.  

\subsection{Phase Conversion and Matched Filtering}
After the channelization, the phase of the complex signals is measured, again in a time multiplexed manner.  The MKIDs in ARCONS have little information in their amplitude response, $(I^2+Q^2)^{1/2}$, so it is ignored~\cite{mazinoe11}.  Given $I$ and $Q$, the phase $\phi = \mbox{tan}^{-1}(Q/I)$ is calculated efficiently on the FPGA using a CORDIC algorithm.

The phase time series is then filtered with a programmable FIR with 26 taps, which gives access to 26$\,\mu$s of data for the filter and allows for pulse arrival time determination to 1$\,\mu$s.  For inputs $x_n$, the FIR is defined conventionally for outputs $y_n$ as
\begin{eqnarray}
y_n = \sum_{m=n-M}^{n}h_m x_m,
\label{FIR definition}
\end{eqnarray}
where $M=25$, and $h_m$ are the FIR taps.  This programmable filter can be loaded with arbitrary coefficients giving wide flexibility.  For single photon detection, we find that a simple 250 kHz low-pass filter, or a matched (Weiner) filter\cite{Turin:1960p5229} provide the best results.  For the matched filter, the coefficients are defined using a template for the expected pulse shape.  In particular, if the normalized pulse template has a profile $p_m$, the coefficients in the above equation are $h_m = p_{(M-m)}$.  This is similar to the technique of pulse compression used in radar and sonar technology\cite{KLAUDER:1960p5383}.  Like pulse compression, it allows for an accurate estimation of the pulse arrival time.  In addition, one can make use of the noise autocorrelation function for a given channel when defining its $p_m$.  When this is done, this filter is often referred to as an ``optimal" linear filter.

\subsection{Event Triggering and Photon Packet Output}
After the data passes the matched filter, it must be analyzed for photon events.  A significant variation of the phase, $\phi$, from the mean indicates the absorption of a photon.  A threshold must be determined for each channel to demarcate incident photons and background electronic noise.  This is done by measuring the mean and standard deviation of each channel while dark, i.e., not illuminated by the sky.  A small memory is allocated on the ROACH board to accumulate a time series of $2^{11}$ samples for a given resonator.  100 snapshots of such data are offloaded to the control PC and used to determine the mean and standard deviation for each channel.  Each threshold is then stored on the FPGA.

The trigger logic is rather simple.  If $\phi$ exceeds the threshold, and $d^2\phi/dt^2 < 0$, the value for $\phi_0$ is recorded as the peak pulse height.  Excessive high frequency noise can cause multiple triggers for a single event.  To avoid this, we implement a 100$\,\mu$s deadtime for each channel, after a pulse has been triggered.  In addition, the two values of the phase before and after $\phi_0$ are recorded.  Recording three values of $\phi$ requires negligible additional bandwidth, but may be valuable for post-acquisition estimation of the pulse height and arrival time.

The arrival time of the photon is also recorded.  A counter runs continuously on each FPGA at 256 MHz.  This counter is reset by the rising edge of each 1 PPS signal from the GPS.  Since the 256 channels are time multiplexed on the FPGA, we are able to resolve the arrival times of photons on each channel to 1$\,\mu$s.

The three values $\phi_{-1}, \phi_0, \phi_{+1}$, the pulse arrival time, and the channel number are then stored in a small memory on the FPGA acting as a packet buffer for offloading the data to the control PC.

\section{Software}
Various software was written to set up the ROACH boards, and for transmitting the acquired data to a control PC.  The setup/control software was written in Python.  In addition, a simple data server written in Python runs on each ROACH board.  This server checks the address pointer of the packet buffer and uploads the packets to the client, the control PC.  At the control PC, the data is accumulated with a c-based client, and is written to disk for later analysis.  Various ``quick-look" tools were written to aid in the acquisition and help with the telescope pointing.

\section{Setup}
A probe signal with large power maximizes the signal and helps quench noise generated by material defects in the resonator~\cite{Gao:2008p66}.  However, the inductance of an MKID is approximately linear only at low probe signal powers.  As the power increases the increasing nonlinear inductance excessively distorts the resonance loop.  We probe each resonator with the maximum power tolerated by the resonator before these effects appear.  

Due to device variations, this maximum power, $P_{max}$ is different for each resonator.  $P_{max}$ generally varies by less than 10 dB between the highest and lowest measured quality factor (Q$_m$) resonators, which can easily be accounted for with the 16 bits of amplitude resolution available to the DAC.  After $P_{max}$ is determined for each resonator, the relative amplitudes of the probe signals are set for the LUT determination.  The output attenuators are then adjusted to set the overall attenuation such that the amplitude of the sum does not exceed the 16 bit range available to the DAC.  

The measured quality factor of the resonators determines the frequency resolution required for the probe signals.  We chose a frequency resolution of 7812.5 Hz and a DAC sample rate of 512 MSample/s.  This means the LUT length must be $2^{16}$ samples of 16 bits, for each $I$ and $Q$.  Each probe signal, $p^{(l)}$, is defined with
\begin{eqnarray*}
p^{(l)}_n &=& I^{(l)}_n + iQ^{(l)}_n \\
		&=& A^{(l)} \mbox{exp}(2\pi if^{(l)} n/512\mbox{e}6 + i\theta^{(l)}),
\end{eqnarray*}
where $n = [0, 1, \ldots, 2^{16}-1]$,  $f^{(l)}$ is the $l^{th}$ resonant frequency, and $\theta^{(l)}$ is a randomly selected phase.  The amplitude $A^{(l)}$ is scaled with respect to the resonator with the highest $P_{max}$, for which $A^{(l)} = 1$.  All $p^{(l)}_n$ are summed and the maximum amplitude of the resultant waveform, $p_{max}$, is determined.  The summed waveform, $p_n$, is then normalized by this maximum amplitude times $2^{16}$ (the largest peak-to-peak amplitude of the DAC) to give
\[
p_n = \frac{2^{16}}{p_{max}}\sum_l p^{(l)}_n,
\]
and casting $p_n$ as a 16 bit integer.  Including the random phase $\theta^{(l)}$ maximizes the dynamic range of the DAC by minimizing $p_{max}$.  This is especially important as the number of resonators, $N$, increases since $p_{max} \rightarrow N^{1/2}$. 
For 256 resonators (the maximum number of channels per board), $p_{max}$ is typically about 50, which results in nearly $-34$ dB power per resonator with respect to the full scale output of a single tone.

With the powers and frequencies set, the next step is to set the threshold separating a pulse due to a photon from noise for each resonator.  The primary consideration for setting this level is the magnitude of the photon pulses we expect to reach our device from the sky through the layers of blocking infrared filters.  A secondary consideration is the data rate limit that data can be pulled from the ROACH boards via the 1 Gb/s ethernet ports.  For ARCONS, we chose between 3 and 4 standard deviations.  Over the course of an observing night, the device temperature slowly drifts up, which requires a re-calibration of the pulse threshold approximately every 4 hours.

The alignment between the ROACH boards and the data acquisition computer is provided by a 1 pulse-per-second signal provided by the GPS on the data acquisition PC.  
	

\begin{figure*}
\includegraphics[width=7.0in]{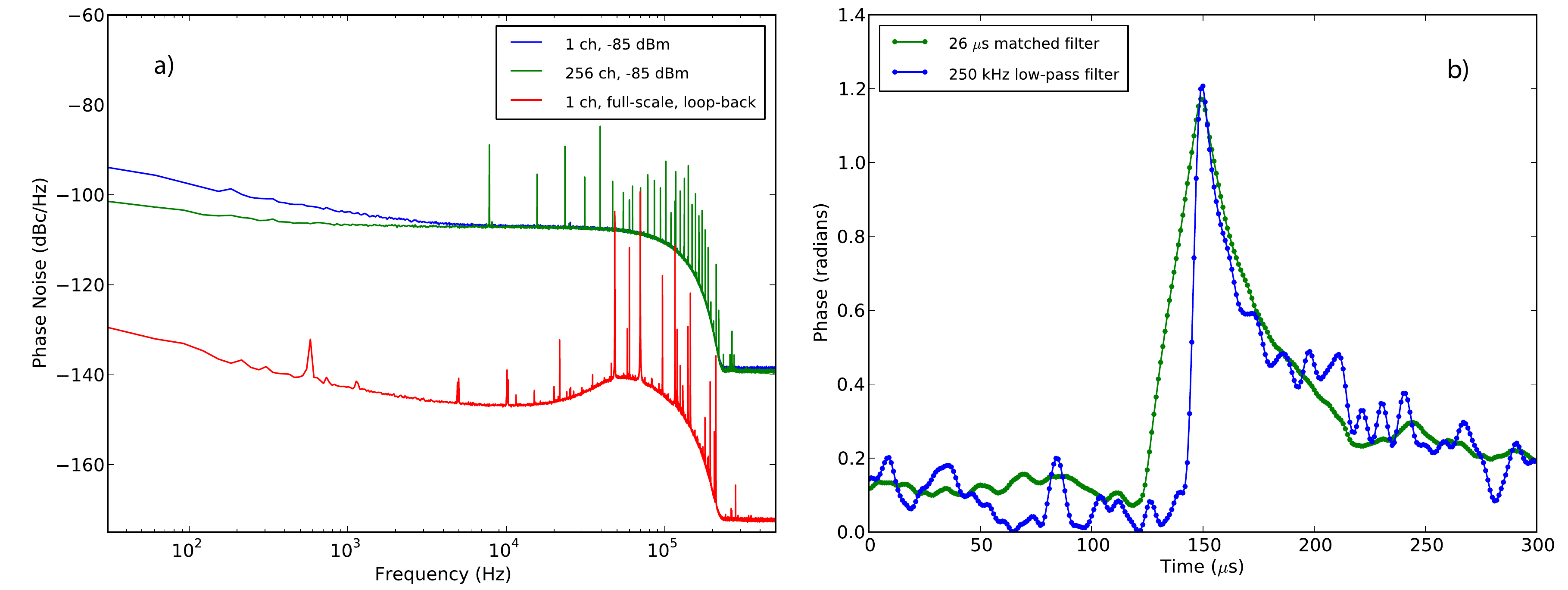}
\caption{(a) Phase noise for the readout under three conditions.  The red line is a loop-back test connecting the IF outputs and inputs, characterizing the noise contributed by the readout electronics.  Aside from the spurious signals, the noise floor is close to the expected -147 dBc/Hz contributed by the ADC.	 The blue and green lines show the phase noise of a single channel through the array with one and 256 probe signals respectively.  Again, the noise floor is close the the expected -106 dBc/Hz contributed by the HEMT amplifier.  (b) Typical phase response of a single channel in time due to 254 nm incident photons for both a low-pass filter and a matched filter.}
\label{fig:results}
\end{figure*}

The performance of the readout is characterized in Figure~\ref{fig:results}.  Since the amplitude of the probe tones are discarded, only a measurement of the phase noise is required.  In order to simplify the comparison with the design requirements listed above, we show phase noise with respect to the origin of the $IQ$ plane and away from the resonance where the transmission through the array is high.  The red line of Fig.~\ref{fig:results}(a) shows the phase noise for a single channel with only one readout tone with the IF out and IF in connected.  This is a measure of the noise introduced by the readout electronics independently of the MKID array, HEMT, and auxiliary room temperature amplifiers.  At mid-range frequencies (100 Hz - 50 kHz) the noise floor is near the expected value of the phase noise contributed by the ADC, -147 dBc/Hz.  There is excess  noise at low frequencies for which we offer no explanation.  At high frequencies, there are spurious tones nearly 40 dB above the noise floor.  The blue line of Fig.~\ref{fig:results}(a) shows the phase noise for a single channel through the array with only one readout tone.  The green line shows the phase noise for the same channel while the full 256 channels are being readout.  With the exception of the spurious tones above 7 kHz, the noise floor of the two are similar and indicate that the readout noise is dominated by the HEMT noise expected to be -106 dBc/Hz, meeting the design requirements.  The spurs appear in harmonics of the frequency resolution of the probe tones (7812.5 Hz).  We therefore speculate that they arise in the IQ mixers as nonlinear mixing products.  

Finally, the ultimate test of our readout scheme is the pulse height estimation, and the energy resolution thereby derived.  In addition to the phase noise, the pulse height estimation is also limited by the length, or number of taps of the programmable stage filter.  To better understand this constraint, we examine the phase signal in time and the response due to incident photons.  Figure~\ref{fig:results}(b) shows the responses to incident 254 nm photons in time using two different FIRs.  The blue curve was taken with the 250 kHz low-pass filter and shows the sudden rise and (roughly) exponential decay to equilibrium, with a lifetime of $\tau \approx$ 50$\,\mu$s.  The green curve was taken with the second stage FIR programmed as a matched filter.  These data show the symmetric response typical of a matched filter.  It also exemplifies the limitations of our FPGA implementation of this filter.  The filter is only able to convolve data over 26$\,\mu$s, which is nearly half of the pulse lifetime, $\tau$.  Effective matched filtering requires that the filter convolution range extend at least $\tau$.  The lack of symmetry in the matched filtered data in Figure~\ref{fig:results}(b) reflects this shortcoming.  Future work will focus on minimizing the channelization logic in order to maximize the length of the programmable filter.  

\section{Summary}
We have demonstrated an effective and scalable system for reading out large arrays of MKIDs using open source hardware and software.  In addition, all firmware and software tools developed here are offered as open-source and downloadable the CASPER website, https://casper.berkeley.edu. 

\begin{acknowledgments}

This material is based upon work supported by the National Aeronautics and Space Administration under a grant issued through the Science Mission Directorate.  The authors would like to thank Jonas Zmuidzinas, Sunil Golwala, Phil Maloney, David Moore, and Danica Marsden for useful insights. A special thanks is extended to Xilinx for donating the FPGAs used in the readout.

\end{acknowledgments}


%


\appendix

\section{Polyphase Filterbank Calculations}

The core of the first stage is formed by a twice over-sampled discrete Fourier transform (DFT) of length $N$.  The DFT is defined  as 
\begin{eqnarray}
\tilde{y}_k = \sum_{n=0}^{N-1}e^{\frac{-2\pi i}{N}nk} y_n,
\end{eqnarray}
where $k = [0, 1, \ldots, N-1]$.  By ``twice over-sampled," we mean the DFT is performed on the overlapping input data sets $[y_0,\ldots, y_{N-1}], [y_{N/2},\ldots, y_{3N/2-1}], [y_N,\ldots, y_{2N-1}], \ldots$, such that each $y_n$ is input twice to consecutive DFTs.  It is instructive to examine consecutive outputs of a bin $k$ with an arbitrary $y_n = e^{2\pi i k^\prime/N}$.  The first output of $\tilde{y_k}$ gives
\begin{eqnarray}
\label{DFT series}
\tilde{y}_k^{(0)} &=& e^{-\pi i n(k-k^\prime)\left(\frac{N-1}{N}\right)}\frac{\mbox{sin}(\pi(k-k^\prime))}{\mbox{sin}(\pi(k-k^\prime)/N)},
\end{eqnarray}
Subsequent outputs accumulate an additional phase forming a time series, $\tilde{y}_k^{(0)}, e^{\pi ik^\prime} y_k^{(0)}, \ldots, e^{\pi imk^\prime} y_k^{(0)}$, with a sample rate reduced by a factor $2/N$.
In general, the input wave frequency $k^\prime$ will be non-integral and $\tilde{y}_k^{(m)}$ will beat accordingly.  For later convenience at the second stage, we introduce an additional factor $(-1)^{mk} = e^{imk\pi}$ to equation \ref{DFT series}.  Now, if our desired resonator frequency is $k^\prime$, we choose bin $k=k^\prime - f$ where $-1\le f < 1$, for finer channelization in the second stage
\begin{eqnarray}
y_k^{(m)} = e^{\pi imf}y_k^{(0)}.
\end{eqnarray}

Equation \ref{DFT series} makes it is clear that the magnitude is strongly attenuated for frequencies $f \ne 0$ with the sinc function that characterizes DFTs.  In order to recover flat frequency response and minimize spectral leakage, the signal must be filtered (Eq. \ref{FIR definition}) before performing the DFT.  Ideally, we would choose the filter such that the sinc function in equation \ref{DFT series} is replaced with a function equal to 1 for $k^\prime$ within the $k^{th}$ bin, and $0$ outside.  The (inverse) Fourier transform of such a rectangular filter in frequency space is a sinc function in the time domain.  In particular, the coefficients for $h_n$ are drawn from a sinc function with $M$ points,
\begin{eqnarray}
h_n = \frac{1}{M}\frac{\mbox{sin}\left[\pi\left(4n/M-1\right)\right]}{\mbox{sin}\left[\frac{\pi}{\alpha M}\left(4n/M-1\right)\right]},
\end{eqnarray}
where $0\le n \le M-1$.  The parameter $\alpha$ allows for the bin-widths to be scaled arbitrarily.  ($\alpha = 1$ gives bin-widths equal to $1/N$.)  Since our objective is to perform a finer channelization on these data for the ultimate 500 kHz narrow bandwidth, we must scale the bins to overlap by at least this amount.  To give sufficient bin overlap, it was chosen as $\alpha = 2.5$.

Since the number of inputs to this filter (convolution) is generally large ($M>N$), the operation is done most efficiently on hardware using the ``Overlap-add" method.  In addition, because $M$ is finite, the behavior of the sinc filter can be improved with a window function, \textit{e.g.}, Hamming, Gaussian, etc.










\end{document}